\documentclass[epj]{webofc}
\usepackage[varg]{txfonts}   % Web of Conferences font
\usepackage{graphicx,amssymb,amsmath,amsfonts,color}
\usepackage[utf8]{inputenc}
\usepackage{subcaption}
\newcommand{\gcmb}{\gamma_{\textsc{cmb}}}

\newcommand{\gcmbebl}{\gamma_{\textsc{cmb},\textsc{ebl}}} %
\woctitle{Quarks-2016. 19th International Seminar on High Energy Physics}
\begin{document}
\title{Constraining Dark Matter and Ultra-High Energy Cosmic Ray Sources with Fermi-LAT Diffuse Gamma Ray Background}

\author{\firstname{Oleg} \lastname{Kalashev}
	\inst{1}\fnsep\thanks{\email{kalashev@inr.ac.ru}} 
%	\and
%        \firstname{Second author} \lastname{Second author}\inst{2}\fnsep\thanks{\email{Mail address for second
%             author if necessary}} \and
%        \firstname{Third author} \lastname{Third author}\inst{3}\fnsep\thanks{\email{Mail address for last
%             author if necessary}}
        % etc.
}

\institute{Institute for Nuclear Research of the Russian Academy of 
	Sciences, Moscow 117312, Russia
%\and
%           the second here 
% \and
%           Last address
          }

\abstract{%
  We use the recent measurement of the isotropic $\gamma$--ray background (IGRB) by Fermi LAT and analysis of the contribution of unresolved point $\gamma$--ray sources to IGRB to build constraints on the models of ultra-high cosmic rays (UHECR) origin. We also calculate the minimal expected diffuse $\gamma$--ray flux produced by UHECR interactions with an interstellar photon background. Finally, for the subclass of dark matter (DM) models with decaying weakly interacting massive particles (WIMP), we build constraints on the particle decay time using minimal expected contributions to the IGRB from unresolved point $\gamma$--ray sources and UHECR.
}
\maketitle
\section{Introduction}
\label{intro}
 The Fermi LAT Collaboration~\cite{Ackermann:2014usa} has presented their measurement of the IGRB within an unprecedentedly wide energy range up to 820 GeV based on 50 months of observations at Galactic latitude ($b$) $|b| > 20$. The IGRB spectrum can be well described over nearly four decades in energy by a power law with exponential cut-off, having a spectral index of 2.3 and a break energy of about 0.3 TeV. The origin of IGRB is not fully understood. However, it has been demonstrated recently~\cite{DiMauro:2016cbj} that unresolved point $\gamma$--ray sources, such as Active Galactic Nuclei or Star Forming Galaxies may account for up to 100\% of IGRB flux and at least 86\% of integral isotropic flux above 50 GeV. This leaves little space for other possible contributions, such as the electromagnetic cascades produced by UHECR interactions and products of Dark Matter (DM) decay (or annihilation). UHECR existence has been confirmed experimentally and therefore the guaranteed minimal contribution to the IGRB from UHECR must exist. On the contrary, the DM-originated flux of $\gamma$--rays is  strongly model dependent and not limited from below. In this work, we discuss both mechanisms in detail. We start from building constraints imposed by the IGRB measurement on UHECR source models in section~\ref{sec-uhecr} and calculate minimal expected cascade $\gamma$--ray flux, originated from UHECR interactions with the interstellar photon background. In section~\ref{sec:DM}, we consider decaying WIMP DM models and use the IGRB spectrum along with minimal contributions to isotropic $\gamma$--ray flux from UHECR and unresolved point sources to build conservative constraints on WIMP decay time, assuming various decay channels. Indirect DM detection methods and their connection to the IGRB have been studied in a number of works (for recent review see~\cite{Gaskins:2016cha}), as well as constraints on UHECR origin models (see e.g. ~\cite{Berezinsky:2016jys} and references therein). To our knowledge however, no study of DM models has been performed taking into account all guaranteed contributions to the IGRB.

\section{UHECR and diffuse $\gamma$--ray background}
\label{sec-uhecr}

Despite the impressive progress in the UHECR experimental study, the chemical composition of cosmic rays at highest energy is still a subject of debate. All modern cosmic ray experiments extract information on the UHECR spectrum and composition from observations of extensive air showers (EAS), initiated by cosmic ray particles. The method involves simulation of EAS development and suffers from unavoidable systematic uncertainties of primary particle energy determination at a level of $15-20\%$ and even larger uncertainties in UHECR composition, since showers initiated by protons and compound nuclei are quite similar. Pierre Auger Observatory (PAO)~\cite{Aab:2014aea}, Telescope Array (TA)~\cite{Abbasi:2014sfa} and  HiRes~\cite{Fedorova:2007cta} observe light element composition in the energy range $1-4EeV$, however PAO reports transfer to heavier composition at higher energies, while TA and  HiRes data is compatible with pure proton composition above $1EeV$.
Observation of accompanying flux of secondary particles from UHECR interactions in intergalactic space can serve as an independent method of the UHECR composition study. In this paper, we focus on the secondary diffuse $\gamma$--ray flux. 
Protons and heavier nuclei produce $e^+e^-$ pairs on the Cosmic Microwave Background (CMB) and infrared-ultraviolet extragalactic background light (EBL) 
\begin{equation} 
N + \gcmbebl \to e^+ + e^- + N. %
\label{PPP}
\end{equation} %
$e^\pm$ then initiate electromagnetic (EM) cascades driven by the chain of inverse Compton scattering of electrons
\begin{equation} %
\label{ICS} %
e^\pm + \gcmb \to e^\pm + \gamma, %
\end{equation} % 
and pair production by high energy photons on the CMB and EBL
\begin{equation} % 
\label{PP} %
\gamma + \gcmbebl \to e^+ + e^- %
\end{equation} %
Relatively fast EM cascade development leading to energy loss and exponential growth of the number of particles proceeds until photons reach the threshold energy for $e^+e^-$--pair production on EBL $E_{th, \gamma}\simeq~\text{TeV}$. The universe becomes essentially transparent for photons with $E_{\gamma} \lesssim 0.1$TeV. For this reason, the energy injected to the EM cascade by UHECR interactions, or by any other mechanisms is accumulated in the form of diffuse $\gamma$-radiation.

The threshold nucleus energy for the process~(\ref{PPP}) on a background photon with energy $\varepsilon$
\begin{equation}
E_{th}=\frac{m_e(m_N+m_e)}{\varepsilon} \simeq 0.5\times A\left(\frac{\varepsilon}{10^{-3}eV}\right)^{-1} EeV,
\label{PPPth}
\end{equation}
is roughly proportional to nucleus mass number $A$. For this reason, compound nuclei are less efficient in $e^\pm$ pair production. Therefore, below we focus on the UHECR source models assuming pure proton composition. We shall see below that IGRB data imposes strict constraints on this class of models.

Also, UHE protons and light nuclei may contribute to the EM cascade through the GZK mechanism~\cite{GZK}
\begin{equation} %
\label{GZK} %
N + \gcmb \to \pi^{\pm,0} + N^\prime, %
\end{equation} %
via products of $\pi^{\pm,0}$ decay, however since the GZK energy threshold for protons is about 50 times higher than $e^{\pm}$--pair production threshold energy~(\ref{PPPth}), its contribution to cascade radiation is 
subdominant unless a very hard primary proton energy spectrum is assumed. 
\subsection{Conservative constraints on proton UHECR sources models}
\label{sec-uhecr-constraint}
Below we use a simple phenomenological model for the UHECR source. We assume homogeneously distributed sources, emitting protons with
power-law generation spectra 
\begin{equation} %
\label{Q_p} %
%\begin{aligned} %
Q_p(E,z) \propto (1+z)^3 H(z) \left(\frac{E}{E_0}\right)^{-p}, \quad
E \in [E_{\min},  E_{\max}], %
%\end{aligned} %
\end{equation} %
where $E_0$ is an arbitrary normalization energy. Below, unless we
state explicitly, we cut the injection spectrum below $E_{\min}=0.1$
EeV and above $E_{\max} = 10^{2.5}$ EeV without loss of generality.
Indeed the main contribution to the EM cascade comes from protons with
energies in the interval from $1$~EeV to a few EeV unless the
injection spectrum is too flat ($\gamma_g \leq 2$) which is forbidden
anyway as it will be demonstrated below. Also note, that
models with $E_{\max}>10^{2.5}$ EeV don't substantially improve UHECR
fit but may overproduce secondary $\nu$-flux. We also introduce the
evolution of source luminosity density with redshift $z$ given by the term $H(z)$
in comoving volume, assuming that the source spectrum shape does not depend on
$z$.
For the evolution term we use general form
\begin{equation} % 
\label{evol} %
H(z)=H(0)(1+z)^m\quad \mbox{for } 0 \leq z \leq z_{\max}. %
\end{equation} %
We also consider two specific cases of source density
proportional to the star formation rate (SFR)~\cite{Yuksel:2008cu}
\begin{equation} %
\label{evol-SFR} %
H_{\rm SFR}(z) \propto \left \{ %
\begin{array}{lll} %
(1+z)^{3.4}, \;\; & z \leq 1 \\ %
(1+z)^{-0.3}, \;\; & 1 < z \leq 4\\ %
(1+z)^{-3.5}, \;\; &z>4. 
\end{array} %
\right. %
\end{equation} %
and evolution of BL Lac/FR I sources~\cite{Giacinti:2015pya}. For the latter, we derived  the following analytical parametrisation
\begin{eqnarray} %
\label{evol-BL} %
H_{\rm BL Lac}(z) \propto  \frac{(a+z)^l}{(b+z)^k}exp(-z/z_c),\quad z<6\\
a=0.000283;\,l=0.850383;\,b=0.7662;\,k=6;\,z_c=10.006 \nonumber
%\right. %
\end{eqnarray} %
%%%%%%%%%%%%%%%%%%%%%%%%%%%%%%%%%%%%%%%%%%%%%%%%
\begin{figure*}[ht!] % 
	\begin{center}
		\begin{minipage}[ht]{68mm}
			\centering
			\vspace{-2mm}
			\includegraphics[width=70 mm, height=50 mm]{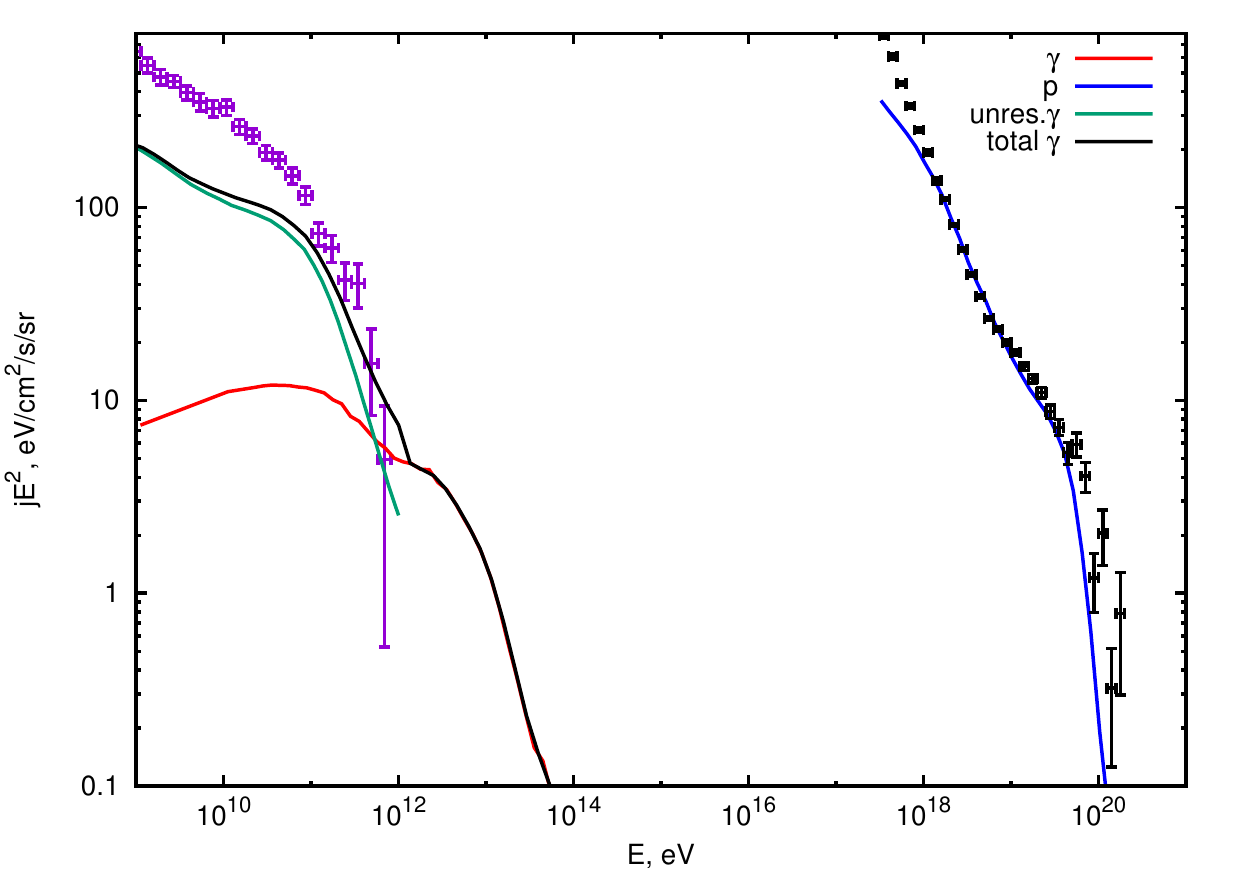}
			\subcaption{$p=2.6$, BL Lac/FR I sources} %
		\end{minipage}
		\hspace{3mm}
		\begin{minipage}[h]{68mm}
			\centering
			\includegraphics[width=70 mm, height=50 mm]{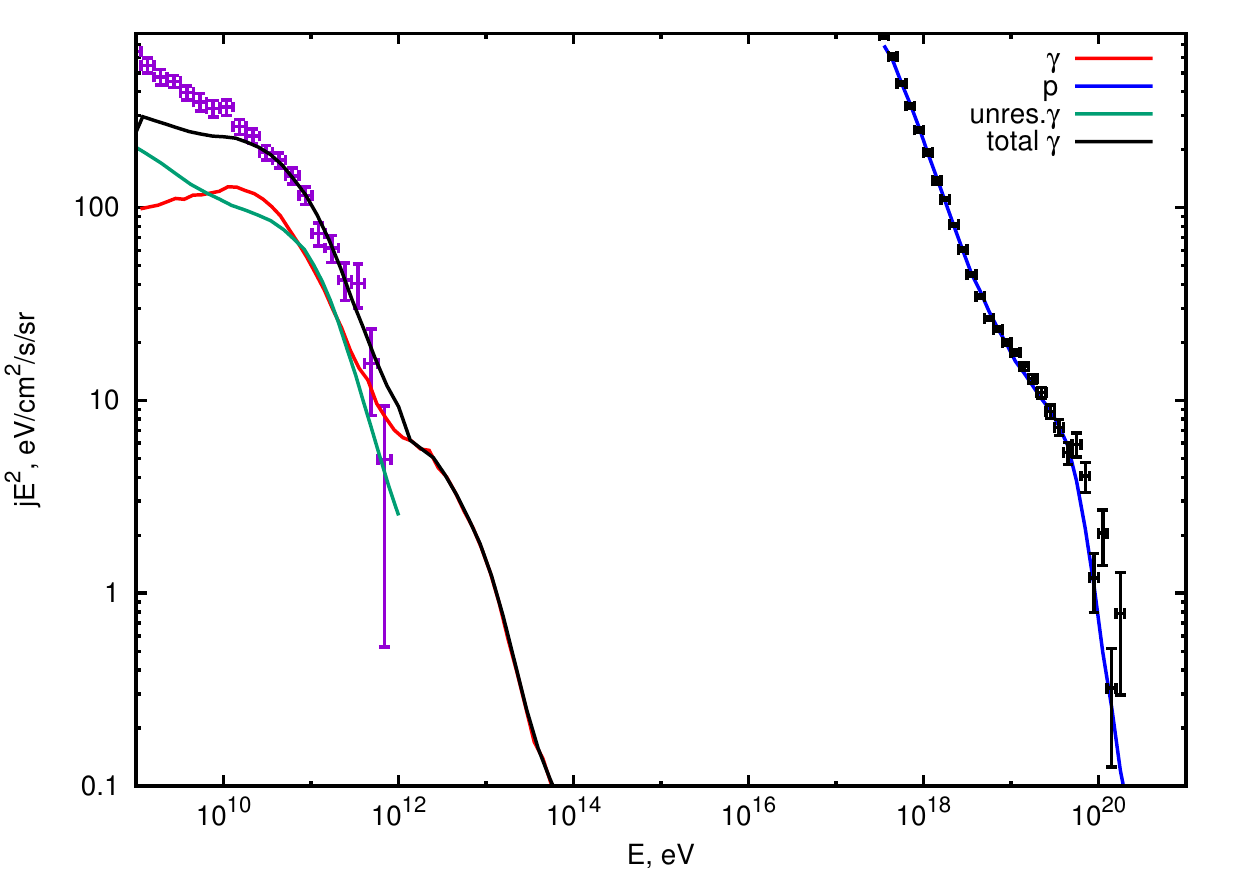}
			\subcaption{$p=2.44$, SFR evolution} %
		\end{minipage}
	\end{center}
	\vspace{-4 mm}%
	\caption{
		The energy spectra of protons (blue curve) and cascade photons (red curve) from sources emitting protons  normalized on the TA spectrum~\cite{TheTelescopeArray:2015mgw}. Also, the Fermi IGRB measurements~\cite{Ackermann:2014usa} are shown assuming minimal galactic foreground allowed by systematic uncertainty, and minimal contribution to IGRB from unresolved $\gamma$--ray sources~\cite{DiMauro:2016cbj} (green curve). Total cascade + minimal unresolved source flux is shown by the black solid curve.
	}
	\label{specSample}
\end{figure*} 
%%%%%%%%%%%%%%%%%%%%%%%%%%%%%%%%%%%%%%%%%%%%%%%%

To obtain the UHECR spectrum and the spectrum of secondary $\gamma$ after propagation through intergalactic space, we use the numerical code~\cite{Kalashev:2014xna} that solves transport equations with terms describing UHECR and electromagnetic cascade interactions with CMB and EBL (for the latter, we use the recent estimate~\cite{Inoue:2012bk}). Following Ref.~\cite{Berezinsky:2016jys}, to derive conservative constraints on proton source models, we don't require a perfect  UHECR spectrum fit in terms of $\chi^2$. Such a requirement would lead to preference for strong evolving models with a relatively hard injection spectrum. These models are strongly constrained by Fermi IGRB and by Ice Cube UHECR $\nu$ flux limit~\cite{Heinze:2015hhp}. Instead, for every $m$ in the physically motivated range
$-1 \leq m \leq 7$ and several values of $z_{\rm max}$ we find optimal value of spectrum power index $p$, providing a reasonable fit. We can do this safely for the following reasons. First of all, the systematic uncertainty of energy spectrum measurement is much higher than the statistical uncertainty in all present experiments. Secondly, the toy effective spectrum model~(\ref{Q_p}) that we use may not precisely describe the actual picture, e.g. the actual spectrum may be different from power law or may be distorted by local sources, etc.

In Fig.~\ref{specSample} we show sample calculations of proton and secondary cascade $\gamma$ made assuming SFR and BL Lac/FR I evolution. One can see (Fig.~\ref{specSample}b) that cascade $\gamma$ flux, when added to the minimal expected contribution from unresolved sources, may easily exceed the IGRB flux measured by Fermi LAT.

To define the formal conservative criterion of model consistency, we use maximal true isotropic flux 
$$\Phi_{\rm max}^{\rm iso}(E_\gamma)\equiv\Phi_{\rm max}^{\rm IGRB}(E_\gamma)-\Phi_{\rm min}^{\rm astro}(E_\gamma),$$
which is calculated by subtraction of minimal unresolved source flux~\cite{DiMauro:2016cbj}   from the maximal IGRB flux allowed by systematic galactic foreground uncertainty.
We compare it with the cascade $\gamma$ flux $\Phi_{\rm cas}(E_i)$ in each IGRB energy bin $i$ with statistical error $\sigma_i$ by constructing $\chi^2_{\rm cas}$:
\begin{equation}
\chi^2_{\rm cas}\equiv\sum_{\Phi_{\rm cas}(E_i)>\Phi_{\rm max}^{\rm iso}(E_i)}\frac{(\Phi_{\rm cas}(E_i)-\Phi_{\rm max}^{\rm iso}(E_i))^2}{\sigma^2_i}
\end{equation}
and require $\chi^2_{\rm cas}<4$ or $\chi^2_{\rm cas}<9$ to obtain an approximate estimate of 2-$\sigma$ or 3-$\sigma$ constraint respectively.
%%%%%%%%%%%%%%%%%%%%%%%%%%%%%%%%%%%%%%%%%%%%%%%%
\begin{figure*}[ht!] % 
	\begin{center}
		\begin{minipage}[ht]{68mm}
			\centering
			\vspace{-2mm}
			\includegraphics[width=70 mm, height=50 mm]{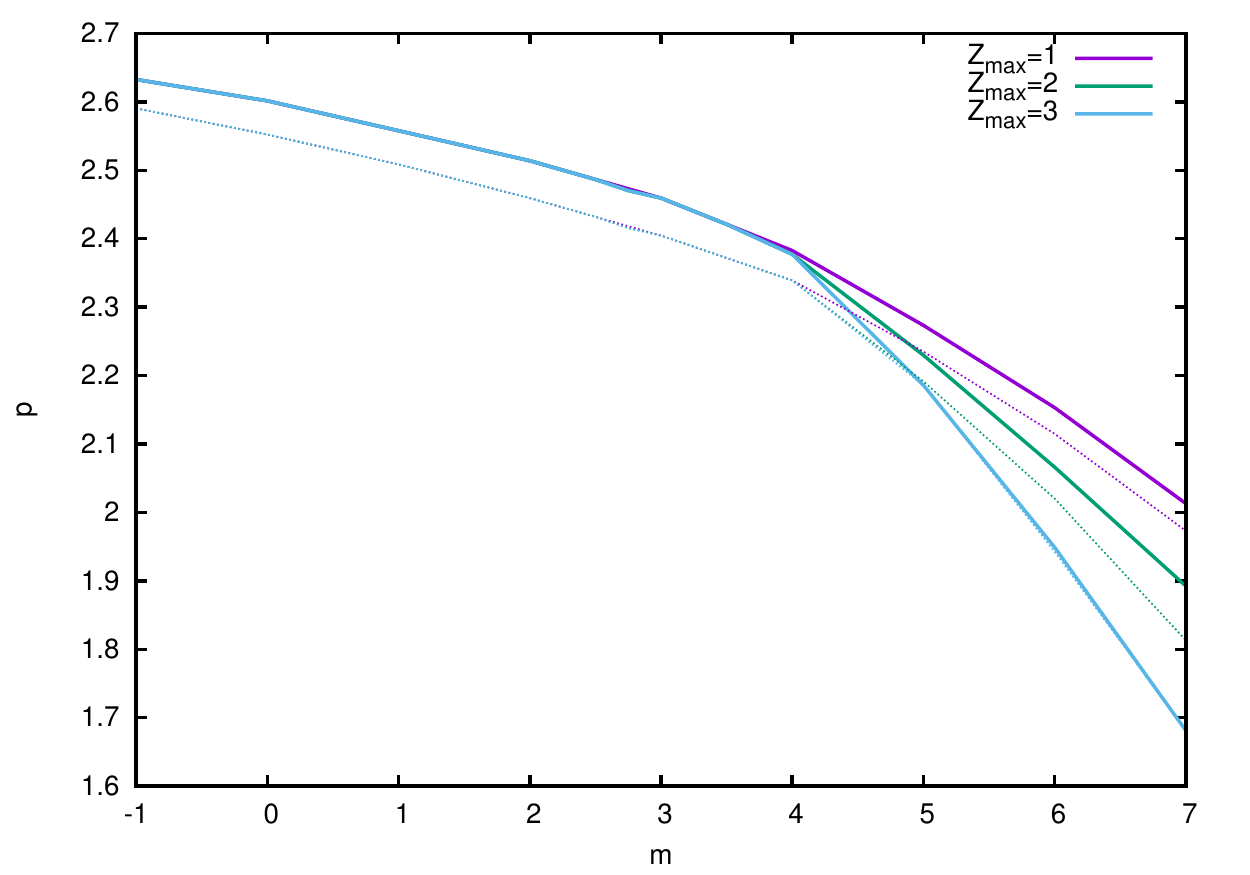}
			\subcaption{best $p(m)$} %
		\end{minipage}
		\hspace{3mm}
		\begin{minipage}[h]{68mm}
			\centering
			\includegraphics[width=70 mm, height=50 mm]{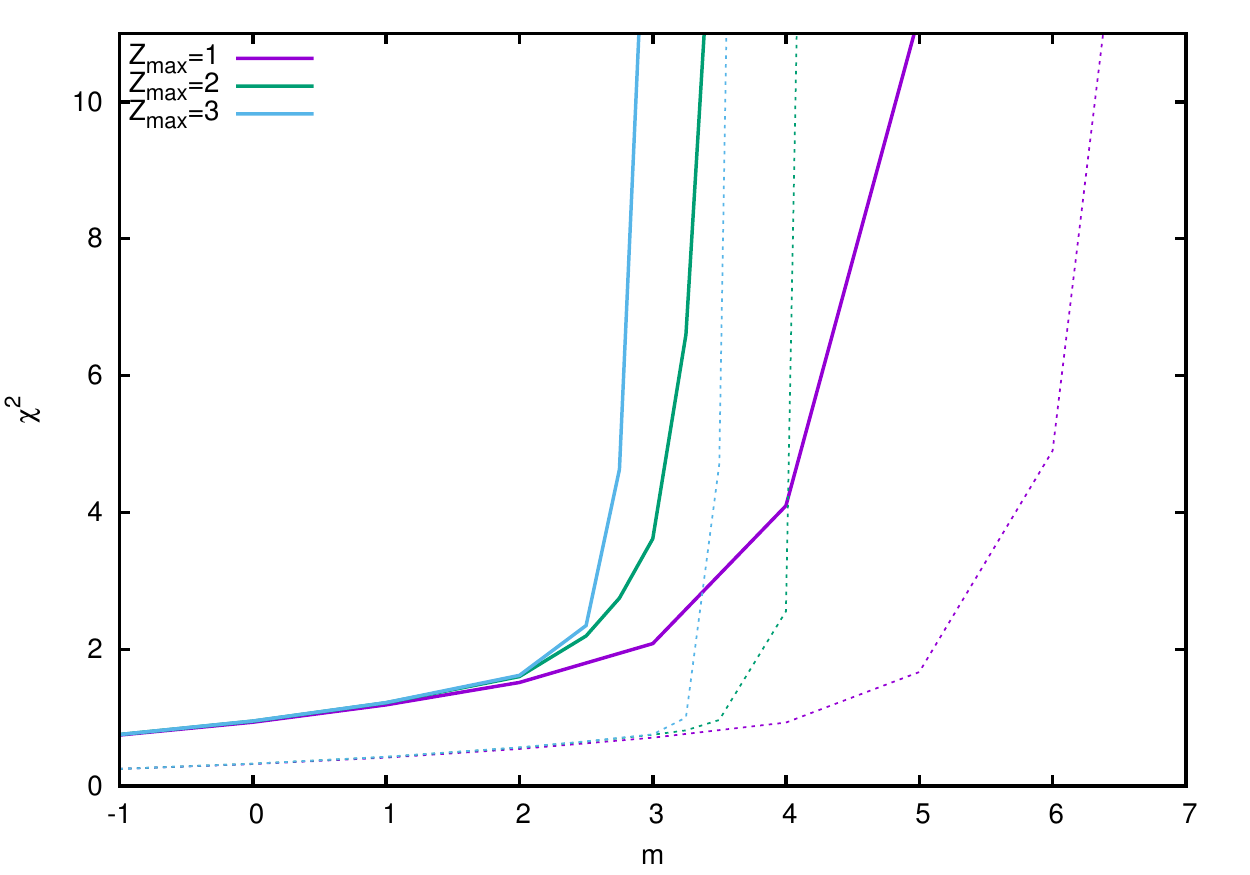}
			\subcaption{$\chi^2_{\rm cas}(m)$} %
		\end{minipage}
	\end{center}
	\vspace{-4 mm}%
	\caption{
		Optimal model parameter $p(m)$ (Fig. a) and $\chi^2_{\rm cas}(m)$ (Fig. b)  obtained for models fitting unshifted (solid lines)  and shifted ($\delta E/E=-20\%$, dotted lines) TA energy spectrum for 3 values of $z_{\rm max}$ 
	}
	\label{fig:chi2}
\end{figure*} 
%%%%%%%%%%%%%%%%%%%%%%%%%%%%%%%%%%%%%%%%%%%%%%%%
The dependence of $\chi^2_{\rm cas}$ on generic model parameters $m$ and $z_{\rm max}$ for models roughly fitting TA energy spectrum is illustrated by Fig.~\ref{fig:chi2}b. One can see that depending on the assumed $z_{\rm zmax}$ models with $m\gtrsim2.5-4$ tend to overproduce diffuse $\gamma$. The constraint weakens if we shift the TA energy scale by 20\% towards lower energies. In this case, models with $m\gtrsim 6$ are rejected for $z_{\max} \geq 1$. Models with stronger evolution would require even smaller $z_{\rm max}$.

\subsection{Minimal diffuse  $\gamma$--ray flux from UHECR sources}
\label{sec-uhecr-min}
As mentioned above, PAO UHECR composition analysis indicates that at least part of UHECR events is caused by compound nuclei. In Ref.~\cite{Aab:2014aea} the fraction of protons was estimated in the total UHECR flux measured by PAO at various energies. To make a realistic estimate of the minimal UHECR contribution  to the diffuse $\gamma$--ray flux, we perform the fit of the proton only component of the PAO UHECR spectrum and calculate the secondary $\gamma$ flux disregarding possible contribution of the heavier nuclei. Also, to minimize the cascade $\gamma$--ray flux, we assume negative though realistic source luminosity density evolution~(\ref{evol-BL}).
%%%%%%%%%%%%%%%%%%%%%%%%%%%%%%%%%%%%%%%%%%%%%%%%
\begin{figure}[h]
	\centering
	\includegraphics[width=0.6\linewidth]{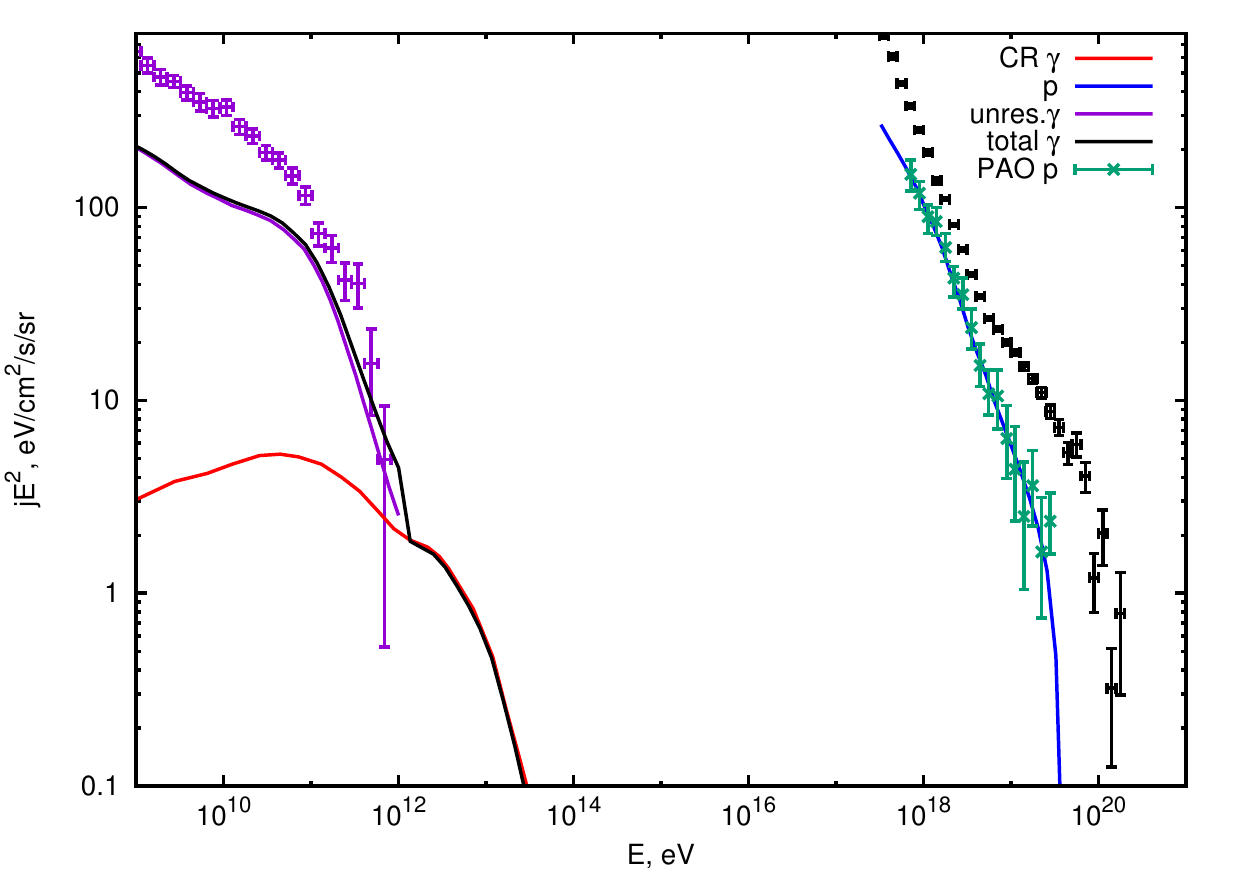}
	\caption{Energy spectra of protons (blue curve) and cascade photons (red curve) from  BL Lac/FR I sources emitting protons with spectrum $E^{-2.79}$, $E<4$EeV normalized on PAO proton spectrum.}
	\label{fig-minUHECR}       % Give a unique label
\end{figure}
%%%%%%%%%%%%%%%%%%%%%%%%%%%%%%%%%%%%%%%%%%%%%%%%
The result of this calculation is presented in Fig.~\ref{fig-minUHECR}. In this case the cascade $\gamma$ flux is well below the IGRB level everywhere except the last energy bin, where it becomes compatible with the flux of unresolved $\gamma$ sources. This result is used in the next section.

\section{WIMP Dark Matter decay and diffuse $\gamma$--ray background}
\label{sec:DM}
In this section, we obtain conservative constraints on WIMP dark matter decay time, based on the observations of the IGRB, taking into account the minimal contribution to the IGRB from unresolved $\gamma$ sources and from UHECR. We use the same method as in the previous section to derive the conservative constraints. Namely, we construct the quantity 
\begin{equation}
\chi^2_{\rm DM}\equiv\sum_{\Phi_{\gamma}^{\rm DM}(E_i)>\Phi_{\rm max}^{\rm DM}(E_i)}\frac{(\Phi_{\gamma}^{\rm DM}(E_i)-\Phi_{\rm max}^{\rm DM}(E_i))^2}{\sigma^2_i},
\end{equation}
where the term $\Phi_{\rm max}^{\rm DM}(E_\gamma)$ stands for the maximal isotropic $\gamma$--ray flux, which could be attributed to dark matter:
$$
\Phi_{\rm max}^{\rm DM}(E_\gamma)\equiv\Phi_{\rm max}^{\rm IGRB}(E_\gamma)-\Phi_{\rm min}^{\rm astro}(E_\gamma)-\Phi_{\rm min}^{\rm cas}(E_\gamma)$$
%%%%%%%%%%%%%%%%%%%%%%%%%%%%%%%%%%%%%%%%%%%%%%%%
\begin{figure*}[ht!] % 
	\begin{center}
		\begin{minipage}[ht]{68mm}
			\centering
			\vspace{-2mm}
			\includegraphics[width=70 mm, height=50 mm]{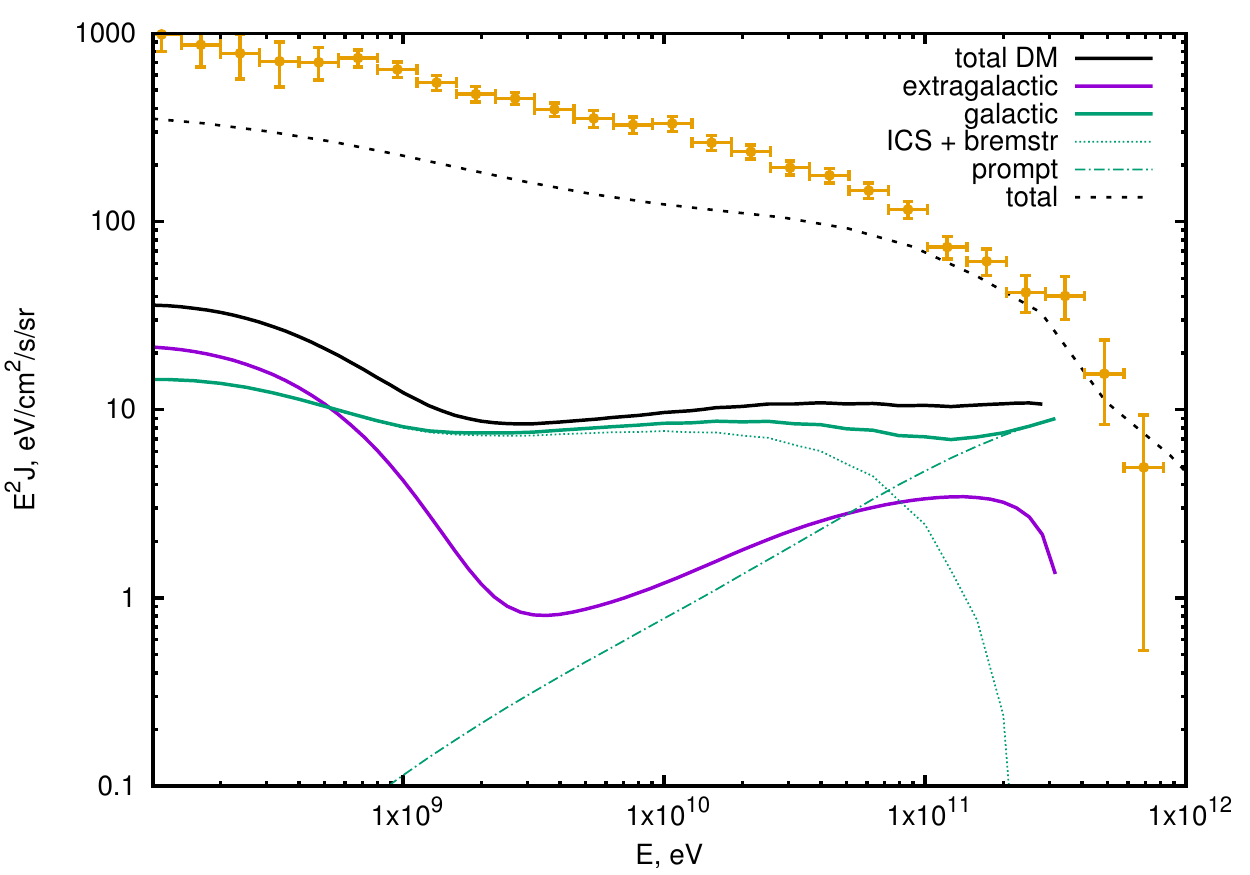}
			\subcaption{$\mu^+\mu^-$ : $\tau=9\times10^{26}s$, $M_{DM}=800GeV$} %
		\end{minipage}
		\hspace{3mm}
		\begin{minipage}[h]{68mm}
			\centering
			\includegraphics[width=70 mm, height=50 mm]{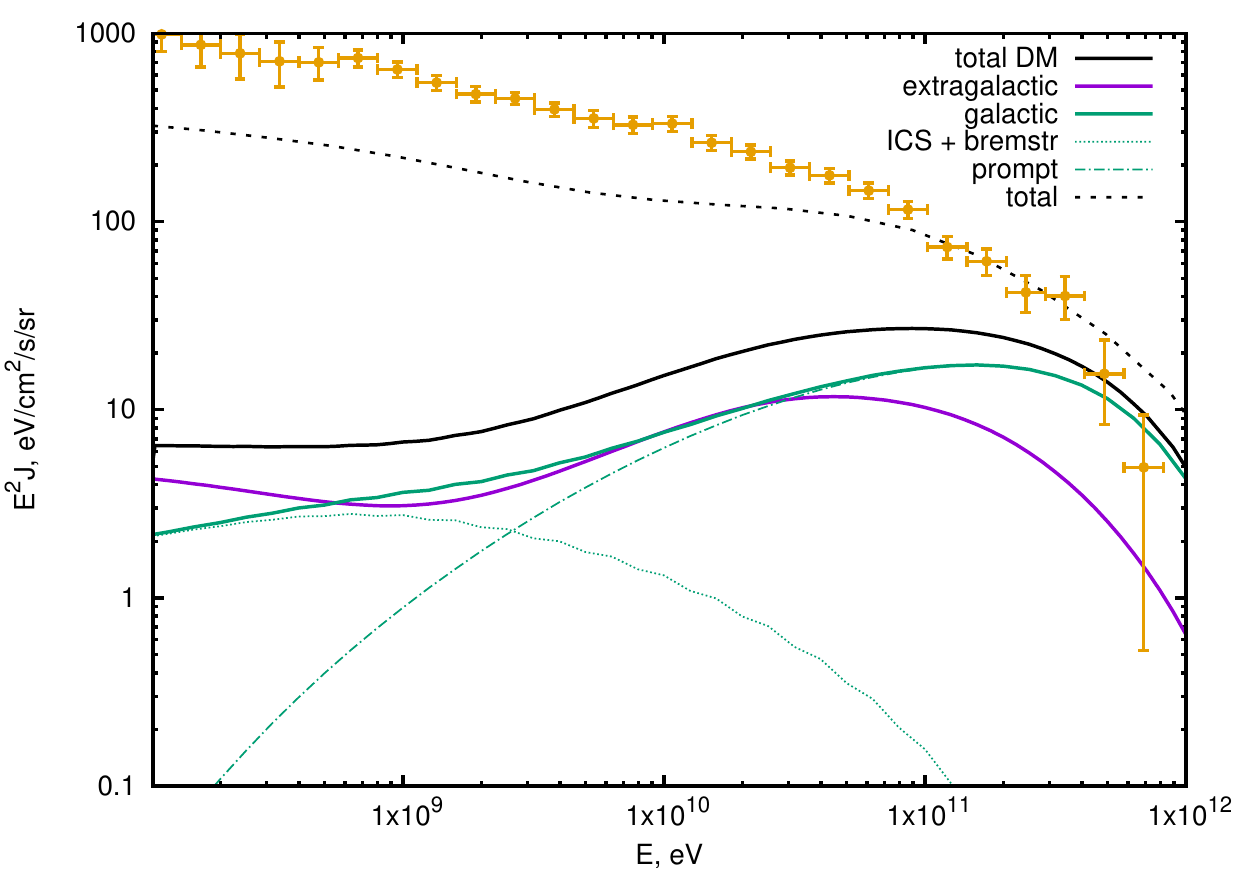}
			\subcaption{u\={u} : $\tau=3\times10^{27}s$, $M_{DM}=5TeV$} %
		\end{minipage}
	\end{center}
	\vspace{-4 mm}%
	\caption{
		Sample spectra of $\gamma$--rays from DM decay.
	}
	\label{fig:DM-sample}
\end{figure*} 
%%%%%%%%%%%%%%%%%%%%%%%%%%%%%%%%%%%%%%%%%%%%%%%%
In fact, the signal from WIMP dark matter is not expected to be isotropic since the DM particles should be concentrated in halos. In our analysis we therefore use the minimal flux, which comes from the direction to the Milky Way anti-centre. We consider decay channels through $e,\mu,\tau,u,b,W$ with  $\gamma$ and $e^\pm$ final states and calculate fluxes of prompt and secondary $\gamma$ from $e^\pm$ interactions. It is produced by the DM in the Milky Way (MW) halo and remote galaxies (EG)
\begin{equation}
\Phi_{\gamma}=\Phi_{\gamma}^{\rm MW} + \Phi_{\gamma}^{\rm EG}
\label{eq:DMtot}
\end{equation}
The prompt galactic $\gamma$--ray flux from the anti-centre direction in the decay channel $f$ with rate $\Gamma_f$ is given by
\begin{equation}
\label{eq:prompt-halo}
\frac{d\Phi_\gamma}{dEd\Omega}=\frac{\Gamma_f}{4\pi M_{\rm DM}\tau_{\rm DM}}  \frac{dN_{\rm \gamma}^f}{dE}  \int_{r_\odot}^\infty \rho(r) dr,
\end{equation}
where $M_{\rm DM}$ and $\tau_{\rm DM}$ are DM particle mass and decay time and  $\rho(r)$ is the DM energy density profile, for which we use the Navarro, Frenk and White parametrisation~\cite{Navarro:1995iw}.
We calculate the injection spectra $dN_{\rm \gamma, e}^f/dE$ and the flux of secondary $\gamma$--rays from bremsstrahlung and inverse Compton scattering of electrons in halo using the public PPPC 4 DM code~\cite{Cirelli:2010xx}. Note that secondary $\gamma$ flux from $e^\pm$ interactions in halo has some uncertainty related to insufficient knowledge of galactic media and magnetic field. We use the ''medium'' interaction model implemented in~\cite{Cirelli:2010xx}.

To calculate the second term in~(\ref{eq:DMtot}) we approximate the extragalactic sources by continuous distribution and simulate the EM cascade propagation taking into account interactions of $e^\pm$ and $\gamma$ with the CMB and EBL using numerical code~\cite{Kalashev:2014xna} with the source density
\begin{equation}
\frac{dQ_{\rm e,\gamma}}{dE}=\frac{\bar{\rho}_0}{M_{\rm DM}\tau_{\rm DM}}(1+z)^3 \Gamma_f  \frac{dN_{\rm e,\gamma}^f}{dE},
\end{equation}
where $\bar{\rho}_0=1.15\times10^{-6} \,{\rm GeV} \, {\rm cm}^{-3}$ is the average dark matter energy density at $z=0$.

In Fig.~\ref{fig:DM-sample} we show the example of these calculations for the decay channels, $M_{\rm DM}$ and $\tau_{\rm DM}$ indicated in captions. Extragalactic and Milky Way halo contributions are shown separately to illustrate that the fluxes are comparable. Note that model shown in Fig.~\ref{fig:DM-sample}a also provides a good fit for AMS-02 positron-electron data~\cite{Liu:2016ngs}.

\begin{figure*}[ht!] % 
	\begin{center}
		\begin{minipage}[ht]{68mm}
			\centering
			\vspace{-2mm}
			\includegraphics[width=70 mm, height=50 mm]{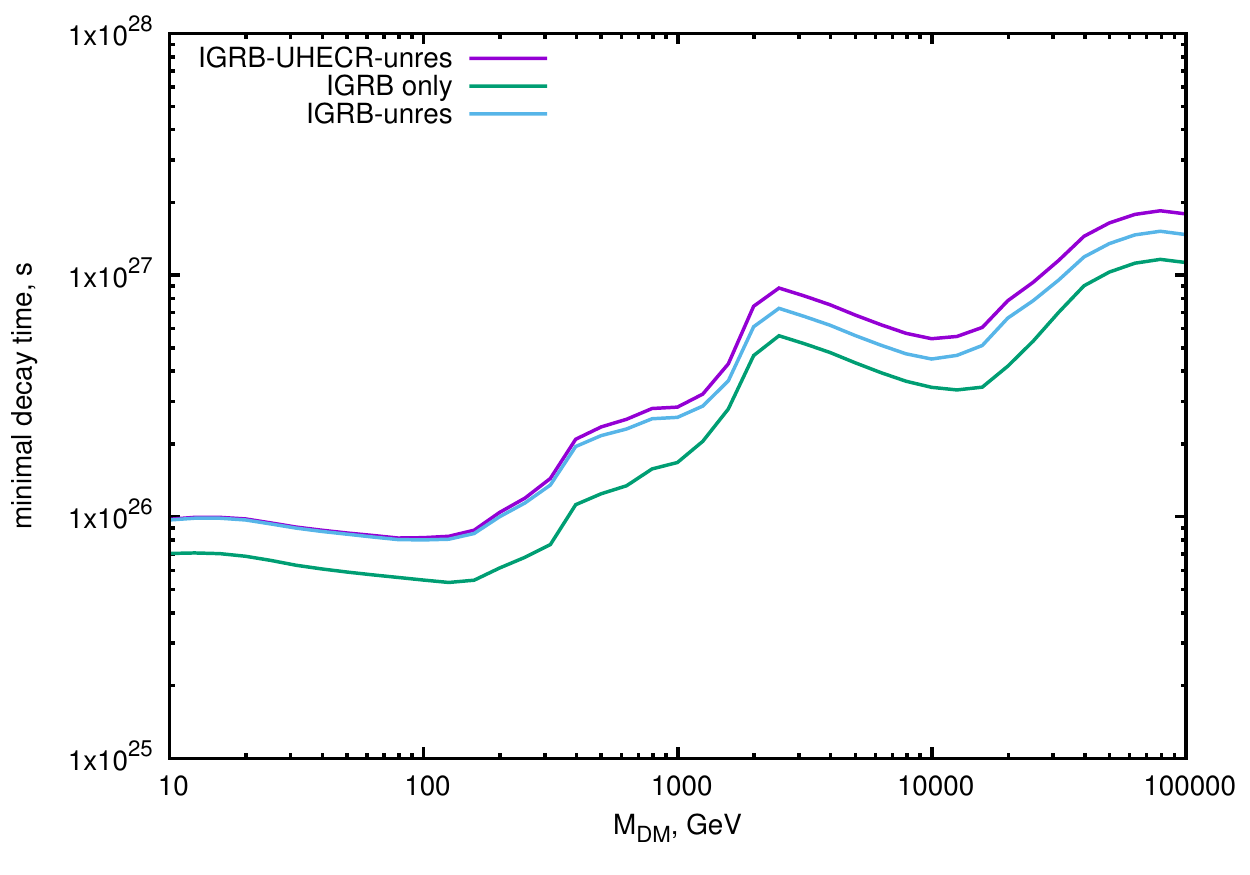}
			\subcaption{$\mu^+\mu^-$} %
		\end{minipage}
		\hspace{3mm}
		\begin{minipage}[h]{68mm}
			\centering
			\includegraphics[width=70 mm, height=50 mm]{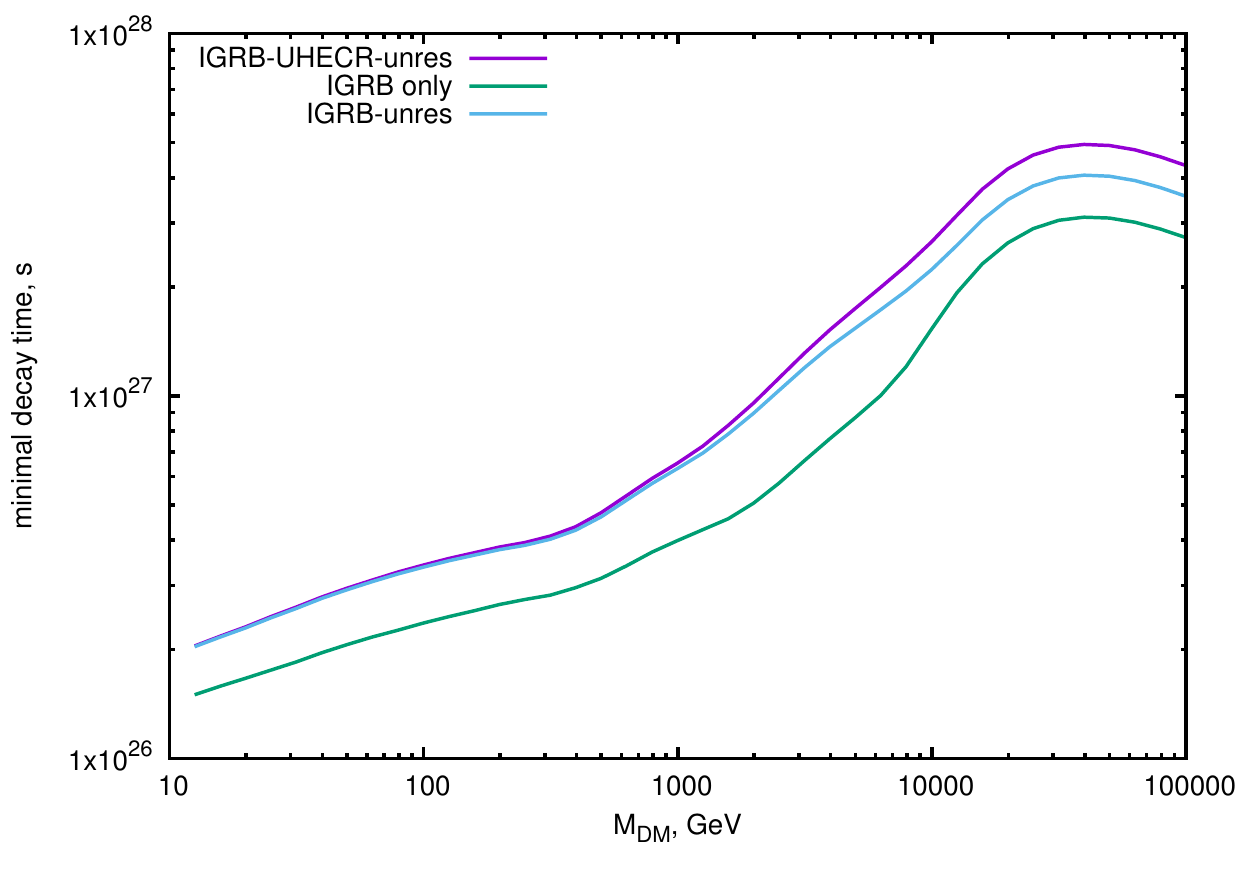}
			\subcaption{b\={b}} %
		\end{minipage}
		\begin{minipage}[ht]{68mm}
				\centering
				\vspace{-2mm}
				\includegraphics[width=70 mm, height=50 mm]{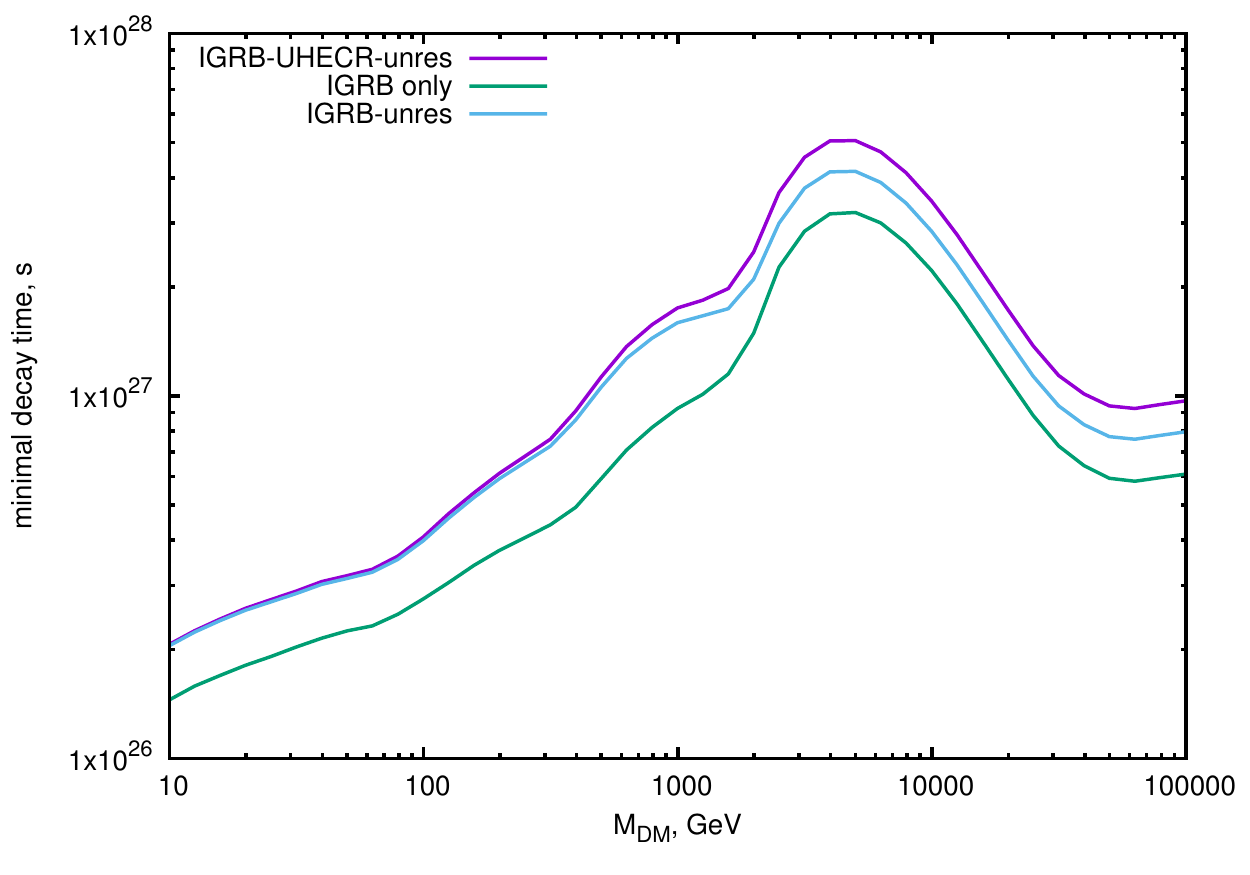}
				\subcaption{$\tau^+\tau^-$} %
		\end{minipage}
		\hspace{3mm}
		\begin{minipage}[h]{68mm}
				\centering
				\includegraphics[width=70 mm, height=50 mm]{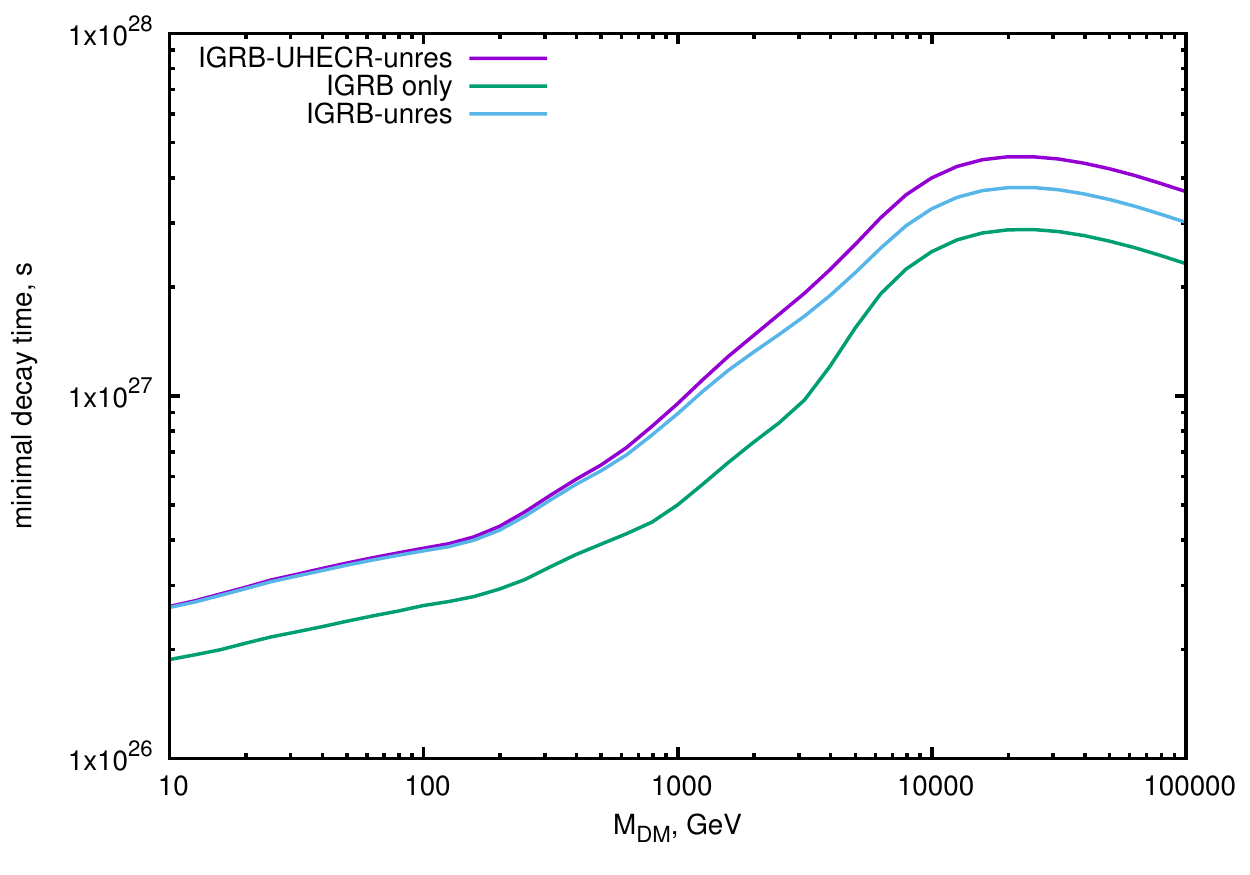}
				\subcaption{u\={u}} %
		\end{minipage}	
		\begin{minipage}[ht]{68mm}
			\centering
			\vspace{-2mm}
			\includegraphics[width=70 mm, height=50 mm]{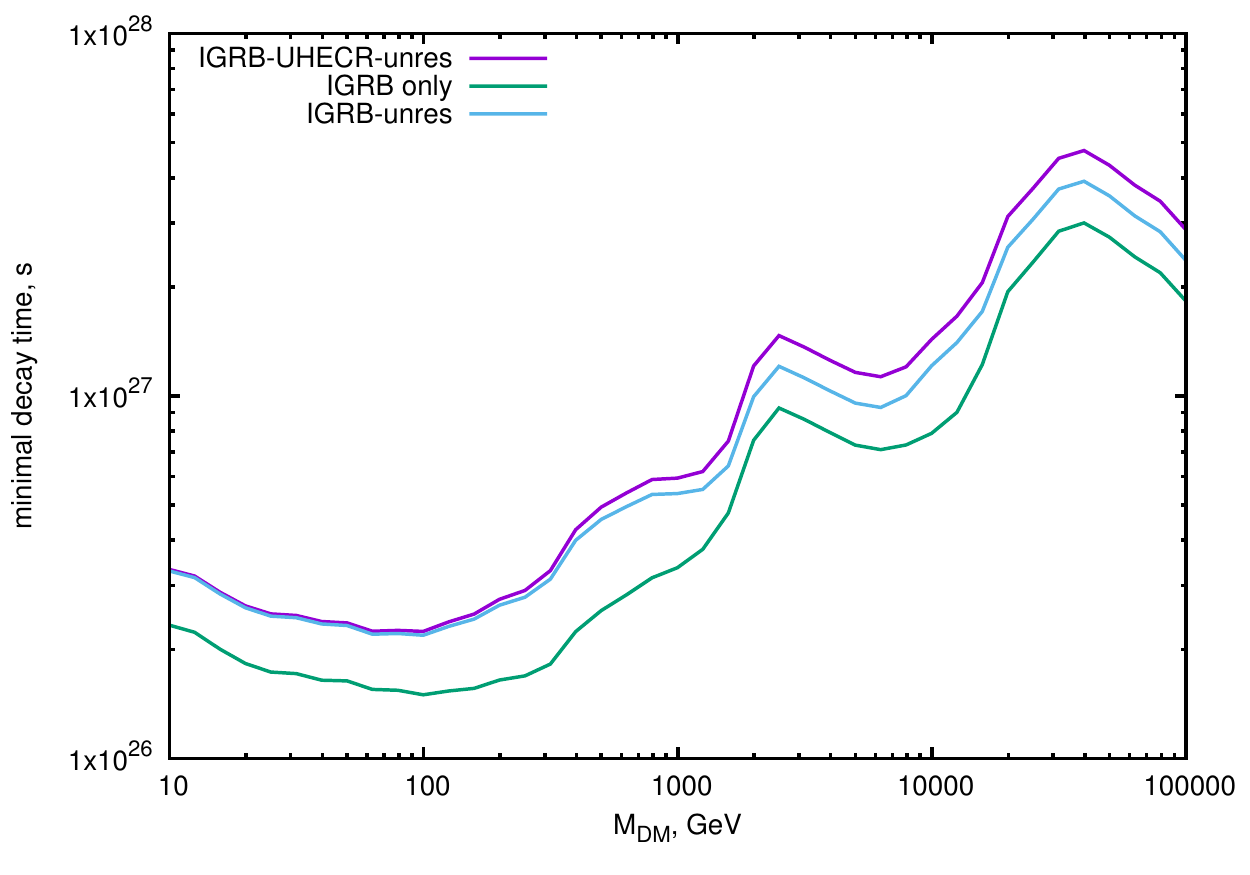}
			\subcaption{$e^+e^-$} %
		\end{minipage}
		\hspace{3mm}
		\begin{minipage}[h]{68mm}
			\centering
			\includegraphics[width=70 mm, height=50 mm]{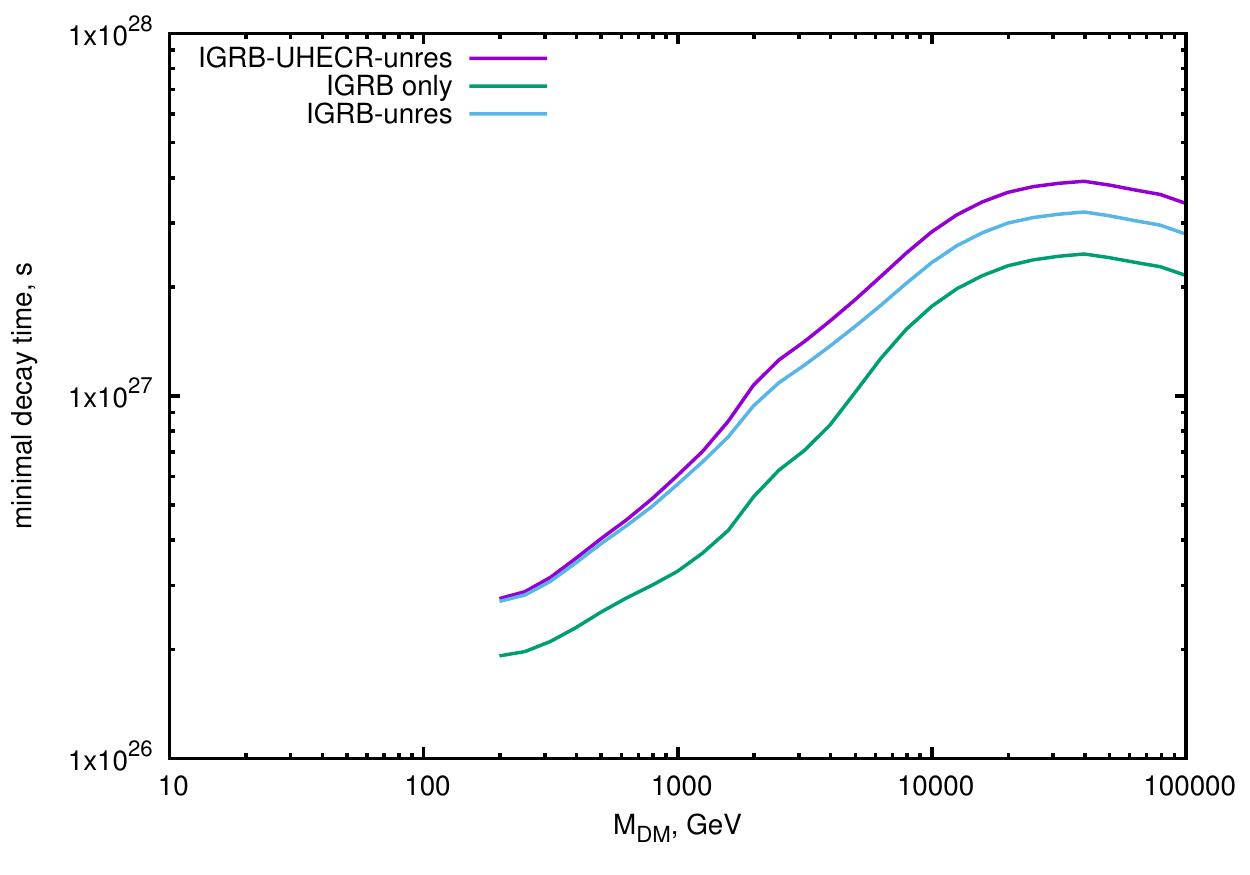}
			\subcaption{$W^+W^-$} %
		\end{minipage}		
	\end{center}
	\vspace{-4 mm}%
	\caption{
		Minimal DM decay time for various decay channels.
	}
	\label{fig:DM-limits}
\end{figure*}

By imposing requirement $\chi^2_{\rm DM}\leq 9$ we constrain the minimal decay time of DM particles in particular channels. These constraints are presented in Fig.~\ref{fig:DM-limits}.
%%%%%%%%%%%%%%%%%%%%%%%%%%%%%%%%%%%%%%%%%%%%%%%%
For comparison, we also show the limits obtained disregarding minimal contribution of UHECR interaction products and unresolved $\gamma$--ray sources (green curves) or disregarding just UHECR products (blue curves). The limits on $\tau_{\rm DM}$ for $M_{\rm DM}\gtrsim \text{few TeV}$ obtained in this work are on average 3-5 times stronger than the conservative constraints presented in the most recent analysis ~\cite{Liu:2016ngs} due to more precise calculation of extragalactic contribution and account for unresolved astrophysical source and UHECR cascade $\gamma$.
%%%%%%%%%%%%%%%%%%%%%%%%%%%%%%%%%%%%%%%%%%%%%%%%
\section*{Acknowledgments}
Section~\ref{sec-uhecr} is a continuation of the original work~\cite{Berezinsky:2016jys} initiated in collaboration with V. Berezinsky and A. Gazizov. The author thanks M. Cirelli for help with the PPPC 4 DM package. The work was supported by the Russian Science Foundation, grant 14-12-01430.  %

%%%%%%%%%%%%%%%%%%%%%%%%%%%%%%%%%%%%%%%%%%%%%%%%

\bibliography{text}

\end{document}